\documentclass[aps]{revtex4}

\usepackage{graphicx}

\begin{document}

\title{Binding energy of shallow donors in a quantum well in the presence of a tilted magnetic field.}

\author{Pawe{\l} Redli\'{n}ski}\email{pawel.redlinski.1@nd.edu}
\affiliation{Physics Department, University of Notre Dame, Notre
Dame, IN 46556}

\author{Boldizs\'{a}r Jank\'{o}}
\affiliation{Physics Department, University of Notre Dame, Notre
Dame, IN 46556}

\begin{abstract}
We present results of variational calculations of the binding energy
of a neutral donor in a quantum well (QW) in the presence of a
magnetic field tilted relative to the QW plane. Assuming that the
donor is located in the center of the QW, we perform calculations
for parameters typical of a \mbox{II-VI} wide-gap semiconductor
heterostructure, using as an example the case of a rectangular
\mbox{CdTe} quantum well with \mbox{CdMgTe} barriers. We present the
dependence of the binding energy of a neutral donor on the tilt
angle and on the magnitude of the applied magnetic filed. As a key
result, we show that measurement of the binding energy of a donor at
two angles of the magnetic field with respect to the quantum well
plane can be used to unambiguously determined the conduction band
offset of the materials building up  heterostructure.
\end{abstract}
\maketitle
%%%%%%%%%%%%%%%%%%%%%%%%%%%%%%%%%%%%%%%%%%%%%%%%%%%%%%%%%%%%%%%%%%%%%%%%%%%
%%%%%%%%%%%%%%%         INTRODUCTION
%%%%%%%%%%%%%%%%%%%%%%%%%%%%%%%%%%%%%%%%%%%%%%%%%%%%%%%%%%%%%%%%%%%%%%%%%%%
\begin{section}{Introduction}
Much work has already been done on calculating the energies and
wavefunctions of electronic states in semiconductor quantum wells
(QWs) in the presence of an applied magnetic field. Most of these
theoretical studies have focused on the binding energy of the
neutral donor ($D^0$) \cite{ref6,ref7,ref8,ref9}, charged donor
\cite{ref6,ref10,ref11}, neutral exciton
\cite{ref12,ref13,ref14,ref15}, and charged exciton (trion)
\cite{ref14,ref15,ref17,ref19,ref20} states as a function of
magnetic field. The vast majority of these calculations treat the
most straightforward case, where the magnetic field is applied
parallel to the direction of growth of the QW. Experimental and
theoretical results show that in this geometry the binding energy of
the above complexes increases with increasing magnetic field
\cite{ref14,ref20}. In spite of considerable progress in this area,
little attention has been paid to the dependence of the binding
energy on the tilt angle of the field relative to the QW plane
\cite{ref1,ref3,ref21}.

The objective of the present work is to determine the dependence of
the binding energy of a neutral donor on the tilt angle $\theta$
between two limiting geometries, the first geometry (denoted below
as Case~\textit{I}) corresponding to the magnetic field $\vec{B}$
aligned along the growth axis of the QW (designated as the
\mbox{z-direction}); and the second limit (denoted as
Case~\textit{II}) corresponding to external magnetic field applied
in the plane of the QW. In our notation described below, in
Case~\textit{I} we define the tilt angle as $\theta=0^{\circ}$, and
in Case~\textit{II} as $\theta=90^{\circ}$, see Fig.~\ref{sketch}
for details.

It is well established that the binding energy  of different
electronic complexes stemming from the Coulomb interaction increases
as the dimensionality of the quantum structure decreases, i.e., as
we progress from quasi-two- to quasi-one-  and eventually to
quasi-zero-dimensional quantum structures \cite{Bastard}. An
external magnetic field localizes the charged particles in the plane
perpendicular to $\vec{B}$ in the form of its cyclotron motion,
while the particle can move freely in the direction of the applied
field, constituting in effect one-dimensional localization
\cite{Yu}. One should note that in this case the density of states
also has the character of a one dimensional system, manifesting
itself as peaks at the Landau level positions. For an electron
subjected simultaneously to the potential of the QW and of an
external magnetic field, 'total' localization of a particle is
different in the two limiting cases defined above. In Case
\textit{I}, the combined action of QW confinement and of magnetic
localization have different directions, which then manifests itself
as quasi-zero-dimensional localization. In Case \textit{II}, the QW
and the magnetic confinements have the same direction, so that the
electron retains its quasi-one-dimensional character associated with
the magnetic confinement. This implies that the binding energy of a
donor should be larger in Case \textit{I} than in Case \textit{II}.
When the tilt angle $\theta$ increases, we can then say that the
dimensionality of an electron is between quasi-zero and quasi-one,
and we expect the binding energy of $D^0$ to be a \textit{monotonic}
function of the tilt angle $\theta$.
\begin{figure}[ht]
  \includegraphics[height=.25\textheight]{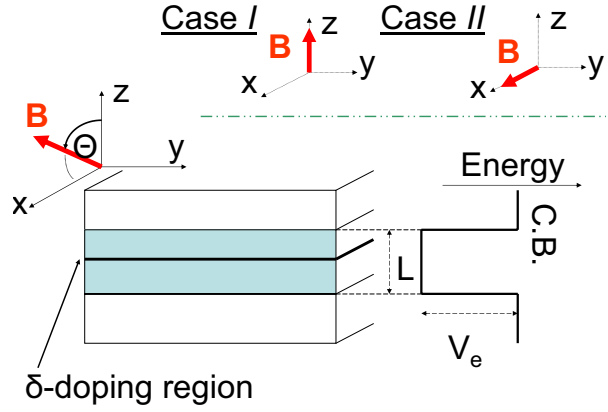}
 \caption{Proposed experimental geometry. CdTe QW is grown in the z-direction:
 $L$ is the QW width and $V_e$ is the QW height.
 The donors are incorporated into the system by \mbox{$\delta$-doping}
 technique only in the center of the well. The edge of the conduction band is
 sketched on the right hand side. The magnetic field $\vec{B}$ lies in
 the XZ plane. Case \textit{I} corresponds to $\vec{B}=B\vec{e}_z$ and Case
\textit{II} corresponds to $\vec{B}=B\vec{e}_x$.}\label{sketch}
\end{figure}
We expect analogous (monotonic) behavior as a function of the tilt
angle for the binding energies of neutral excitons (composed of two
charged particles) or trions (composed of three charged particles).
These predictions are in contradiction to the results obtained for
the binding energy of neutral donors in a \mbox{GaAs} quantum well
presented in Ref.~\cite{ref21}.

The present paper is organized as follows. We start with  a model
Hamiltonian and a class of trial wave functions. We then present
results of variational calculations for the \mbox{CdTe}~QW with
\mbox{CdMgTe} barriers. These include calculations of the binding
energy $E_b$ as a function of the tilt angle $\theta$, as a function
of the magnetic field $B$, and as a function of the quantum well
barrier height $V_e$. We show, finally, that measurement of the
binding energy of $D^0$ in the two limiting geometries
(Case~\textit{I} and Case~\textit{II} on Fig.~\ref{sketch}) can be
used as a tool for determining the conduction band offset at the
well/barrier interface.
\end{section}
%%%%%%%%%%%%%%%%%%%%%%%%%%%%%%%%%%%%%%%%%%%%%%%%%%%%%%%%%%%%%%%%%%%%%%%%%%%
%%%%%%%%%%%%%%%         THEORY
%%%%%%%%%%%%%%%%%%%%%%%%%%%%%%%%%%%%%%%%%%%%%%%%%%%%%%%%%%%%%%%%%%%%%%%%%%%
\begin{section}{Model}
The Hamiltonian of a shallow donor embedded in
a symmetric square quantum well is modeled by the following
Hamiltonian
\begin{equation}\label{ham}
    H=\frac{(\vec{p}-e\vec{A}(\vec{r}))^2}{2m_e^*} + V^{QW}(z) -
\frac{e^2}{4\pi\epsilon\epsilon_0}\frac{1}{|\vec{r}-\vec{R}_0|}-\mu_Bg_e^*\vec{B}\,\vec{s}.
\end{equation}
The first part of the Hamiltonian is the kinetic energy of a
delocalized conduction electron (where $e$ is the electron charge,
and $m_e^*$ is its effective mass) in presence of a tilted magnetic
field \mbox{$\vec{B}=B(\sin(\theta),0,\cos(\theta))$} lying in the
XZ plane, see Fig~\ref{sketch}. We have chosen an asymmetric gauge
for vector potential \mbox{$\vec{A}(\vec{r})=B(0,
x\cos(\theta)-z\sin(\theta), 0)$}. For $\theta=0^{\circ}$ (Case
\textit{I}), the magnetic field $\vec{B}$ is parallel to the OZ
axis; and for $\theta=90^{\circ}$ (Case \textit{II}) it lies in the
XY plane (i. e., in the plane of the QW), see Fig~\ref{sketch}. The
profile of the potential energy of the QW is described by the second
term in Eq.~(\ref{ham}):
\begin{equation}\label{profile}
V^{QW}(z)=\Big\lbrace \begin{array}{cc}
              0 & |z| > L/2 \\
              -V_e & |z| < L/2
              \end{array},
\end{equation}
where $L$ is the width of the QW centered at $z=0$, and $V_e$ is the
height of its barrier. The energy scale is chosen by defining the
conduction band edge of the barriers as zero. The third term in
Eq.~(\ref{ham}) is the Coulomb energy of a shallow donor located at
point $\vec{R}_0$. We assumed that the donor center is located at
the center of the QW, so we can set $\vec{R}_0 = \vec{0}$ without
losing physical generality. The last expression in Eq.~(\ref{ham})
is the Zeeman Hamiltonian, in which $g_e^*$ is the effective
\mbox{g-factor} of conduction electrons. We find it inconvenient to
rotate the Hamiltonian, Eq.~(\ref{ham}), to a new coordinate system
in which the kinetic energy has a simpler form than in the original
system, because then the functional form of the potential profile,
Eq.~(\ref{profile}), couples two variables ($x$ and $z$ in our
notation) in a non-trivial way \cite{ref1,ref4}.

So far the problem of the Hamiltonian of a donor in a QW has not
been solved analytically (even the case of a free electron in a QW
in a tilted magnetic field remains analytically unsolved
\cite{ref2,ref4,ref5}), so that in our work we have used a
variational approach. We propose the following form of the trial
wave function of a \mbox{two-component} spinor,
\begin{equation}\label{trialWF}
\Psi_{\pm}=f_{\pm}(\vec{r})\cdot\chi_{\pm}=\sum_{i=0}^{N_1}\sum_{j=0}^{N_2}\sum_{k=0}^{N_3}
C^{\pm}_{ijk} \phi_i(\alpha x)\phi_j(\alpha y)\phi_k(\alpha z)\cdot
\chi_{\pm},
\end{equation}
where  $\phi_i$ are a one-dimensional harmonic oscillator functions
(Gaussian functions) and $\chi_{\pm}$ are spin states. Note that,
Gaussian trial wave functions have been successfully used in
Ref.~\cite{ref6,ref15} for calculating donor and trion states,
respectively. The nonlinear variational parameter $\alpha$ (the
scaling parameter), and the linear variational parameters
$C^{\pm}_{ijk}$ were determined using the Ritz variational method.
In Eq.~(\ref{trialWF}) the number of the basis functions has to be
finite, and in this connection we have checked that $N_1 = N_2 = N_3
= 10$ are sufficient to ensure that the results do not depend on the
cut-off of the number of basis functions.

The orbital part of the total wave function of a donor $\Psi_{\pm}$,
Eq.~(\ref{trialWF}), is denoted by $f_{\pm}(\vec{r})$, $\chi_{\pm}$
being the spin part. It is easy to show by direct substitution that
the two spinors, \mbox{$\chi^{\dagger}=\left(cos(\theta/2),
sin(\theta/2)\right)$} and
\mbox{$\chi^{\dagger}=\left(-sin(\theta/2), cos(\theta/2)\right)$},
solve the Schr\"{o}dinger equation that contains the Hamiltonian
given by Eq.~(\ref{ham}). Then the orbital part $f_{\pm}(\vec{r})$
of the spinor function $\Psi_{\pm}$ satisfies the following
eigen-equation:
\begin{widetext}
\begin{equation}\label{EEfpm}
\{\frac{(\vec{p}-e\vec{A}(\vec{r}))^2}{2m_e^*} + V^{QW}(z) -
\frac{e^2}{4\pi\epsilon\epsilon_0}\frac{1}{|\vec{r}|} \mp
\frac{1}{2}\mu_Bg_e^*B\}f_{\pm}(\vec{r})=E_{\pm}f_{\pm}(\vec{r}).
\end{equation}
\end{widetext}
Additionally, as seen from the above equation, $f_{\pm}(\vec{r})$
has the same functional form for both spin configurations
$\chi_{\pm}$.
\end{section}
%%%%%%%%%%%%%%%%%%%%%%%%%%%%%%%%%%%%%%%%%%%%%%%%%%%%%%%%%%%%%%%%%%%%%%%%%%%
%%%%%%%%%%%%%%%         RESULTS AND DISCUSSION
%%%%%%%%%%%%%%%%%%%%%%%%%%%%%%%%%%%%%%%%%%%%%%%%%%%%%%%%%%%%%%%%%%%%%%%%%%%
\begin{section}{Results and Discussion}
In order to demonstrate the tilt angle dependence of the binding
energy of $D^0$, we performed calculations for two barriers heights,
\mbox{$V_e$=200~meV} and 20~meV. These two choices of barriers
height correspond to 19\% and 2\% content of Manganese in the
barriers, respectively. Additionally, for each $V_e$ we chose
$L$=100~\AA~and $L$=300~\AA; and \mbox{$B$=0, 1, 4, 9, and 16~T},
corresponding to magnetic lengths \mbox{$\lambda_c$=$\infty$, 256,
130, 85, and 65~\AA}. For specificity we used CdTe materials
parameters: \mbox{$m_e^*$=0.1} of free electron mass for both  the
QW and the barrier, and dielectric constant $\epsilon$=10.4, which
give the characteristic Coulomb scales: \mbox{$Ry^*$=12.6~meV} and
\mbox{$a^*_B$=55~\AA}.

The donor binding energy $E_b$ is obtained as a difference between
the ground state energies of the free electron and of the donor
($E_{\pm}$) with the same electron spin configuration \cite{Yu}.
This definition implies that the Zeeman Hamiltonian does not
contribute to the binding energy of $D^0$.
\begin{figure}[ht]
  \includegraphics[height=.3\textheight]{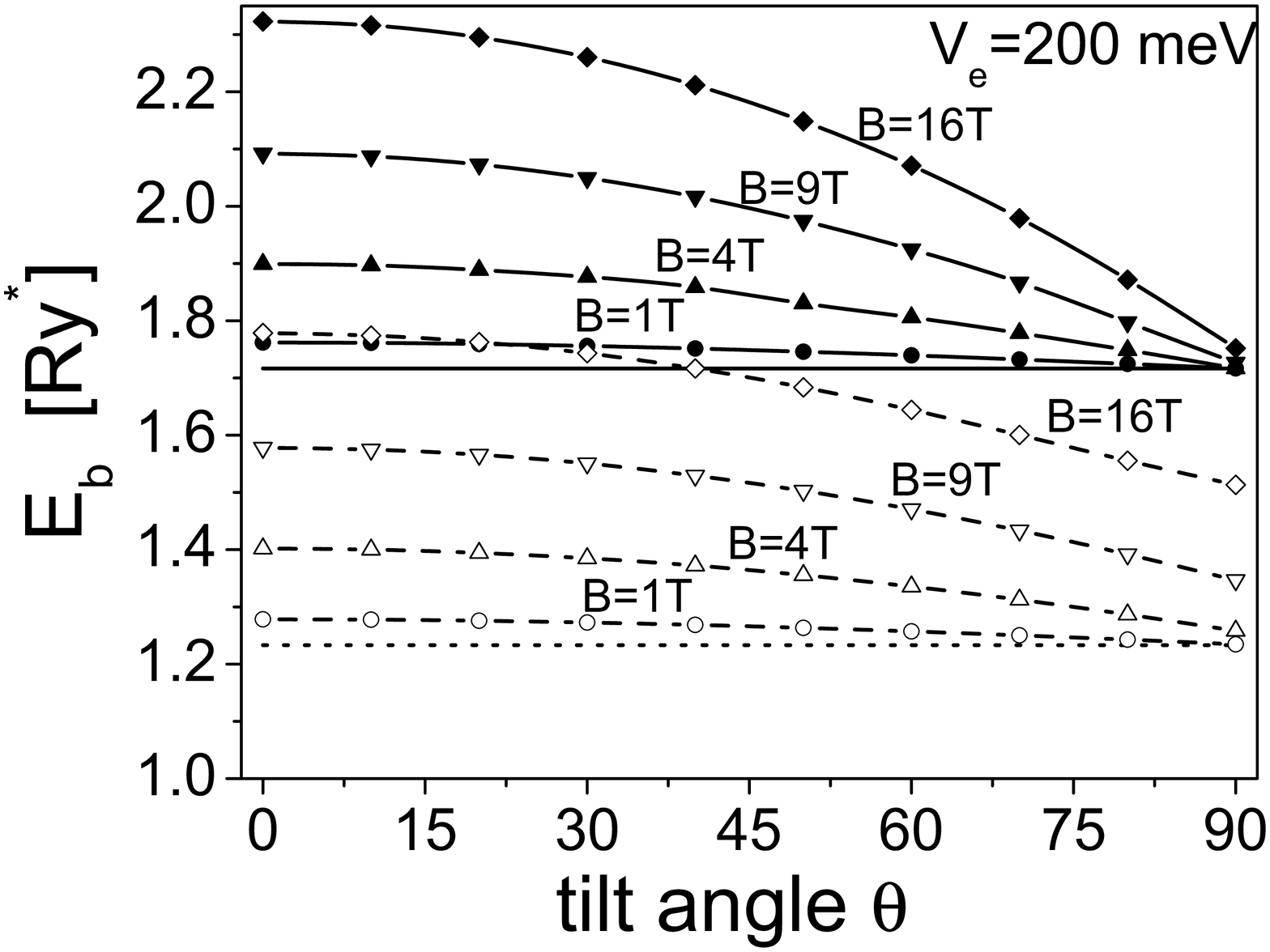}
 \caption{Donor binding energy $E_b$ in a rectangular \mbox{CdTe/Cd$_{0.81}$Mg$_{0.19}$Te} quantum well
 as a function of tilt angle $\theta$ of external magnetic field $B$ for different magnetic field values
 (different symbols), calculated for two QW widths $L$. Full lines correspond to \mbox{$L$=100~\AA},
 and dashed lines to \mbox{$L$=300~\AA}. Lines without symbols correspond to \mbox{$B$=0~T}.}\label{fig1}
\end{figure}
We will be interested mainly in the variation of $E_b$ with the
inclination $\theta$ and the magnitude of the magnetic field,
$E_b=E_b(\theta, B)$. It must be noted that $E_b$ also depends on
other parameters, e. g., the barrier height, but these are kept
constant in each variational process.

%First we show the results obtained for different molar fraction of
%Mg in the barrier: x=19\% and x=5\%, which correspond to
%$V_e\approx$200~meV and $V_e\approx$50~meV, respectively.

In Fig.~\ref{fig1} we present the binding energy of the donor ground
state as a function of tilt angle $\theta$ for $V_e$=200~meV at
different magnetic fields, as well as for two different QW widths
$L$=100 and 300~\AA. First we discuss the $L$=100~\AA~case,
represented by full lines in Fig.~\ref{fig1}.

At $B$=0~T, the binding energy of the neutral donor is about
1.7~$Ry^*$=21.4~meV, and is increased by 70\% compared to the
binding energy of a donor in three dimensions (in this case $L \sim
a_B$). Next, for $B \ne 0$ T, the binding energy is a
\textit{monotonically} decreasing function of the tilt angle: at a
given magnetic field \mbox{$B$ = const}, the binding energy is
highest for $\theta=0^{\circ}$, and decreases for
$\theta>0^{\circ}$. Suspection of Fig.~\ref{fig1} shows that for
$\theta=0^{\circ}$ the difference $E_b(B=16T)-E_b(B=0T)$ is
0.6~$Ry^*$=8~meV, while for $\theta=90^{\circ}$ it is only
0.05~$Ry^*$=0.6~meV, see also Fig.~(\ref{figVsB}). These totaly
different values in the two limiting field orientations (case
\textit{I} and case \textit{II}) are related to the fact that the QW
width is smaller then (or comparable to) the characteristic magnetic
length $\lambda_c$ at fields up to 16~T. If the magnetic field is
applied along the z-direction ($\theta=0^{\circ}$), the electron is
localized in x- and y-directions by the external magnetic field, as
discussed at the outset (the bigger the field, the larger the
magnetic localization). When this effect is combined with QW
confinement, the electron becomes localized in all three direction.
As the magnetic field is changing from 0~T to 16~T, the initially
quasi-two-dimensional electron is becoming increasingly
quasi-zero-dimensional. We thus expect that the binding energy will
increase substantially in this situation. On the other hand, if the
magnetic field is aligned in the x-direction, magnetic localization
involves the y- and z-directions with characteristic lengths
$\lambda_c$. Since the QW also confines the electron motion in the
z-direction, and does not restrict its motion along x, the combined
effects of magnetic and of QW localization now result in a
one-dimensional motion. As seen in Fig.~\ref{fig1}, even up to
\mbox{$B$=16~T} the magnetic localization now has practically no
effect, and binding energy practically does not depend on $B$.
\begin{figure}[ht]
  \includegraphics[height=.3\textheight]{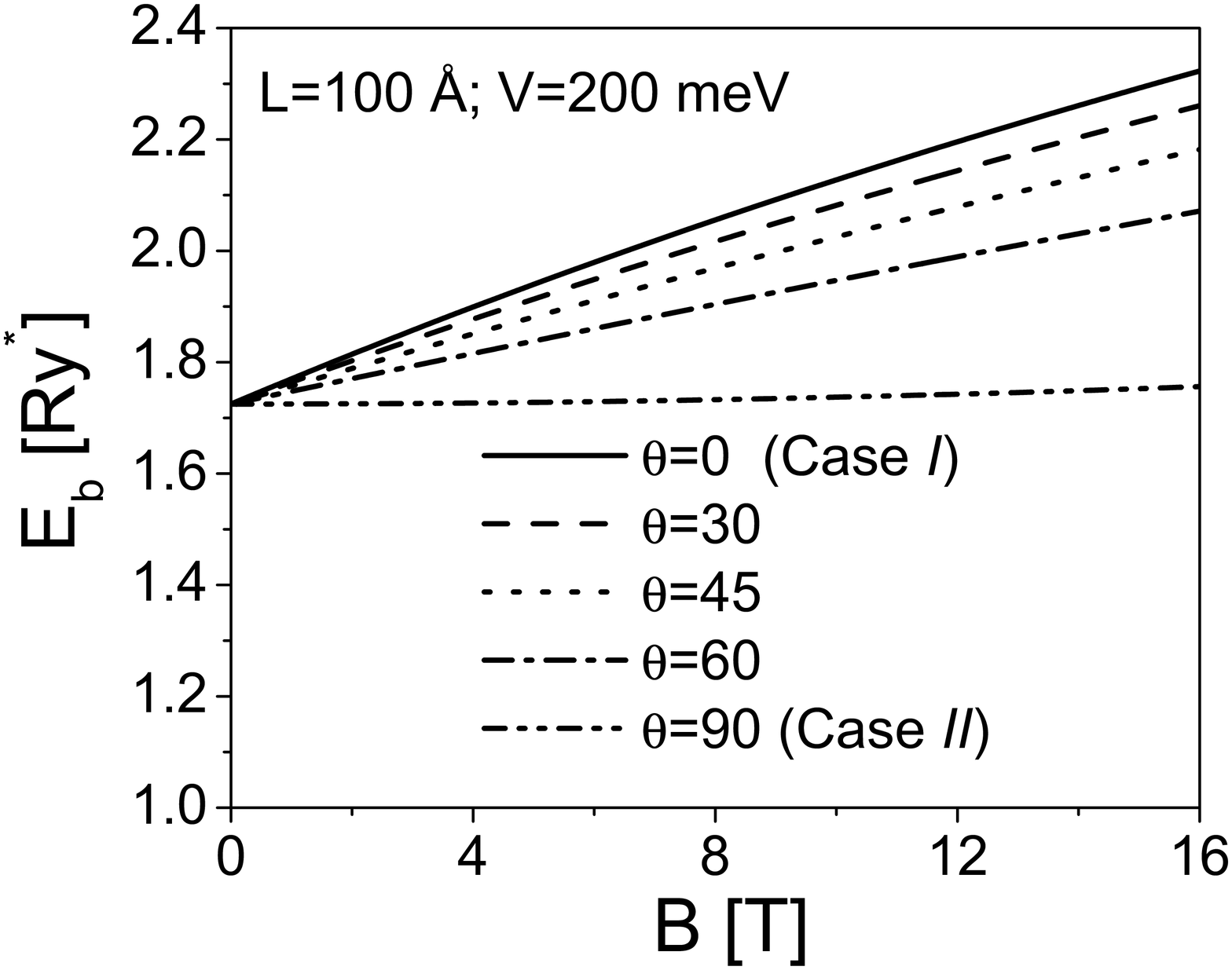}
 \caption{Donor binding energy $E_b$ in a \mbox{CdTe/Cd$_{0.81}$Mg$_{0.19}$Te} QW
 ($L$=100~\AA) as a function of a magnetic fields $B$ for $\theta$=0$^{\circ}$ (Case \textit{I}),
 30$^{\circ}$, 45$^{\circ}$, 60$^{\circ}$ and 90$^{\circ}$ (Case \textit{II}). In the Case \textit{II},
 the binding energy is practically independent on the magnetic field $B$ ($B<$16~T).}\label{figVsB}
\end{figure}
In contrast with $L$=100~\AA, for a wider QW ($L$=300~\AA) the
changes in binding energy produced by the magnetic field are quite
visible at $\theta$=90$^{\circ}$, as seen in Fig.~\ref{fig1}. For
such a wide QW, the change of $E_b$ as $B$ increases from 0 to 16~T
is 0.3~$Ry^*$=4~meV. Now $L$=300~\AA~and the binding energy is only
1.2~$Ry^*$ (without magnetic field) so that the system is nearly
three-dimensional ($a_B<<L$), in contrast to the two-dimensional
character obtained for $L$=100~\AA. Thus for all values of $\theta$
the magnetic field effectively localizes the particle in both
dimensions perpendicular to the direction of the applied field. This
explains why the curves for $E_b$ are much more flat for
$L$=300~\AA~than for $L$=100~\AA, particularly at higher values of
$B$. We expect that, when we increase the QW width even more, $E_b$
should become even more flat, eventually approaching the
\mbox{three-dimensional} limit, where it ceases to depend on the
tilt angle even for large $B$. Our calculations clearly confirm this
trend.

In Fig.~\ref{figVsB} we show the dependance of the donor binding
energy $E_b$ for a \mbox{CdTe/Cd$_{0.81}$Mg$_{0.19}$Te} quantum well
($L$=100~\AA) as a function of  the magnetic field $B$, for five
different values of the tilt angle $\theta$. We have chosen the
magnetic field range to be $0 \leq B \leq 16 \, {\rm T}$, which is
the most widely accessible field range in photoluminescence (PL)
spectroscopy. As seen in the figure, the binding energy at $B$=0~T
is $E_b (B \sim 0 T) \simeq 1.7 \, {\rm Ry}^*$, which corresponds to
a 70\% increase in binding energy with respect to the binding energy
of a three-dimensional donor. This increase is due to the
confinement generated by the quantum well. For Case \textit{I}, the
increase in magnetic field has a clearly visible impact on $E_b
(B)$.  In contrast, $E_b$ is practically constant for Case
\textit{II}. While the series of curves presented in
Fig.~\ref{figVsB} seem to be linear, our results for the binding
energy have, in fact, a square-root dependence on the external
magnetic field. Such a dependence is in accordance with the results
obtained in Ref.~\cite{ref6}. The apparent linearity of the curves
is due to the fact that at the highest field we consider ($B$=16~T)
the ratio $\gamma
\def \frac{\hbar \omega_{c}}{2 Ry^*}$ is only 0.74. The $
E_b (B) \propto \sqrt{B}$ scaling behavior becomes apparent for a
much wider range of magnetic fields $ 0 \leq \gamma \leq 5$.
\begin{figure}[ht]
  \includegraphics[height=.3\textheight]{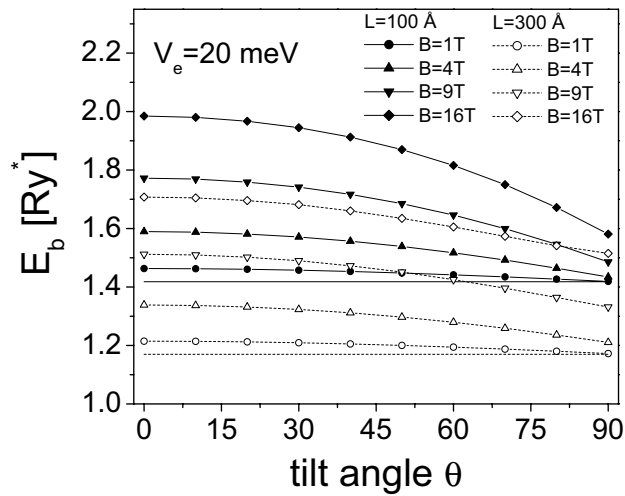}
 \caption{Donor binding energy $E_b$ in a rectangular \mbox{CdTe/Cd$_{0.98}$Mg$_{0.02}$Te} quantum well
 as a function of tilt angle $\theta$ of external magnetic field $B$ for different magnetic field values
 (different symbols), calculated for two QW widths $L$. Full lines correspond to \mbox{$L$=100~\AA},
 and dashed lines to \mbox{$L$=300~\AA}. Lines without symbols correspond to \mbox{$B$=0~T}.}\label{fig2}
\end{figure}
In Fig.~\ref{fig2} we present results for $V_e$=20~meV
(corresponding to $x \approx 0.02$). Comparing Fig.~\ref{fig1} and
Fig.~\ref{fig2}, we see that the binding energy of the donor is
larger for $V_e$=200~meV than for $V_e$=20~meV. This well known fact
originates from the larger quantum confinement of the deeper QW. The
characteristics of the results in Fig.~\ref{fig2} are similar to
those in Fig.~\ref{fig1}, including the \textit{monotonic} behavior
of $E_b$. Comparing the curves in Fig.~\ref{fig1} with corresponding
curves in Fig.~\ref{fig2}, we see that the latter clearly are more
flat. This again confirms that in the three-dimensional case, i.e.,
as $V_e\rightarrow 0$, we should have no $\theta$ dependance
(straight horizontal lines).
\begin{figure}[ht]
  \includegraphics[height=.3\textheight]{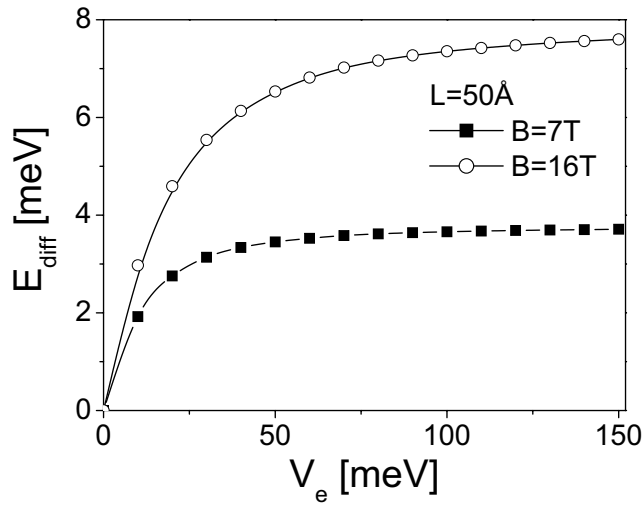}
 \caption{Difference $E_{diff}$ of the binding energy of a donor for $\theta$=0 and
 $\theta$=90$^{\circ}$ in a quantum well with $L = 50$ \AA\, as a function of barrier height $V_e$. Full symbols correspond to
$B$=7~T, and open symbols
 to $B$=16~T.}\label{fig3}
\end{figure}
In Fig.~\ref{fig3} we show the difference  $E_{diff}$ between the
binding energy of a donor at $\theta=0^{\circ}$ and its binding
energy at  $\theta=90^{\circ}$ as a function of the height of the
barrier $V_e$; i.e.,
$E_{diff}=E(\theta=0^{\circ})-E(\theta=90^{\circ})$. At $B$=7~T and
$V_{e}>$~50~meV, the difference $E_{diff}$ is practically constant
and relatively small (only 3.7~meV), but at $B$=16~T it does not
saturate until \mbox{$V_e\approx$75~meV}; and its value is twice as
high, i.e. 7.5~meV. This feature can be utilized as a tool for
determining the conduction band offset, at least in QWs with
moderate barrier heights \cite{Kossut1}. Note that the bigger the
magnetic field, the larger the offset which can be measured using
this method.

In Ref.~\cite{ref21}, Fig.~6, the binding energy of the donor as a
function of tilt angle is a non-monotonic function showing a maximum
at $\theta=45^{\circ}$, in contrast to the monotonic behavior
reported here. In our opinion this is related to the approach
employed by the authors of Ref.~\cite{ref21}, in which a real
quantum well is transformed into two QWs oriented at right angles to
one another. In Ref.~\cite{ref1} the same group, using the same
approximation, calculated the exciton binding energy as a function
of tilt angle (see Fig.~7 in Ref.~\cite{ref1}). Unfortunately, the
approach used in the latter reference does not provide the results
for the range of $\theta$ between $0^{\circ}$ and $15^{\circ}$, and
between $75^{\circ}$ and  $90^{\circ}$, which appears to be an
artefact of the technique used in Ref.~\cite{ref1} and
Ref.~\cite{ref21}.
\end{section}
%%%%%%%%%%%%%%%%%%%%%%%%%%%%%%%%%%%%%%%%%%%%%%%%%%%%%%%%%%%%%%%%%%%%%%%%%%%
%%%%%%%%%%%%%%%         CONCLUSIONS
%%%%%%%%%%%%%%%%%%%%%%%%%%%%%%%%%%%%%%%%%%%%%%%%%%%%%%%%%%%%%%%%%%%%%%%%%%%
\begin{section}{Conclusions}
We have shown the results of variational
calculations of the binding energy of a neutral donor in a
rectangular quantum well as a function of the angle of en external
magnetic field tilted with respect to the growth direction of the
QW. In our calculations we used parameters characteristic of
\mbox{II-VI} compounds (using specifically parameters for the
\mbox{CdTe/CdMgTe} QW system), and we assumed that the donor is
located in the center of the quantum well. For a given magnetic
field, the largest binding energy is found to correspond to the case
when the magnetic field is perpendicular to the plane of the QW. We
find that the binding energy of $D^0$ is a \textit{monotonic}
function of the tilt angle $\theta$, decreasing with increasing tilt
angle, in contrast with earlier calculations reported in
Refs.~[\cite{ref1,ref21}]. Our results reduce to the
three-dimensional limit when either the quantum well width increases
or the barrier height decreases, providing a "reality check" for the
method used. We have shown that for the
\mbox{CdTe/Cd$_{1-x}$Mg$_x$Te} quantum well ($x < 0.1$), the
conduction band offset can be determined by measuring the binding
energy of the neutral donor at two perpendicular directions of the
applied magnetic field, $\theta$=0 and 90$^{\circ}$. To our
knowledge this technique of determining conduction band offsets has
not been previously recognized.
\end{section}
%%%%%%%%%%%%%%%%%%%%%%%%%%%%%%%%%%%%%%%%%%%%%%%%%%%%%%%%%%%%%%%%%%%%%%%%%%%
%%%%%%%%%%%%%%%         ACKNOWLEDGEMENT
%%%%%%%%%%%%%%%%%%%%%%%%%%%%%%%%%%%%%%%%%%%%%%%%%%%%%%%%%%%%%%%%%%%%%%%%%%%
\begin{section}{Acknowledgement}
We would like to thank P. Bogus{\l}awski, J. Kossut and J. K.
Furdyna for their interest in this subject and useful discussions.

This research was supported in part by the National Science
Foundation under NSF-NIRT Grant No. DMR 02-10519, by the U.S.
Department of Energy, Basic Energy Sciences, under Contract No.
W-7405-ENG-36 and by the Alfred P. Sloand Foundation (B. J.).
\end{section}

\end{document}